\documentclass[conference]{IEEEtran}
\usepackage{amsmath,amsthm}
\usepackage{subfigure}
\usepackage{cite}
\usepackage{graphicx}
\usepackage{epstopdf}
\usepackage{color}
\usepackage{amsfonts,amsmath,amssymb}
\usepackage{cite}
\usepackage{graphicx}
\usepackage{url}
\usepackage{bm}
\usepackage{bbm}
\usepackage{amssymb}
\usepackage{braket}
\usepackage[T1]{fontenc}
\usepackage{fancyhdr}
\begin{document}
	
	\title{\huge{Multi-mode CV-QKD with Non-Gaussian Operations}}
	\author{
		\thanks{Financial support provided by Northrop Grumman Corporation. Mingjian He is partially supported by the China Scholarship Council.}
		\IEEEauthorblockN{Mingjian He$^1$,  Robert Malaney$^1$, and Jonathan Green$^2$}\\
		
		\IEEEauthorblockA{$^1$School of Electrical Engineering  \& Telecommunications,\\
			The University of New South Wales,
			Sydney, NSW 2052, Australia. \\
			$^2$Northrop Grumman Corporation, San Diego,
			California, USA. }}
	
	\maketitle
	\thispagestyle{fancy}
	\renewcommand{\headrulewidth}{0pt}

	
\begin{abstract}
	Non-Gaussian operations have been studied intensively in recent years due to their ability to increase the secret key rate for certain CV-QKD protocols.
	However, most previous studies on such protocols are carried out in a single-mode setting, even though in reality any quantum state contains multi-mode components in frequency space.	
In this work we investigate the use of non-Gaussian operations in a multi-mode CV-QKD system.
Our main finding is that, contrary to single-mode CV-QKD systems, in generic multi-mode CV-QKD systems it is possible to use non-Gaussian operations to increase the optimized secret key rate. More specifically, we find that at losses of order 30dB, which represents a distance of order 160km and the effective maximum distance for CV-QKD, the key rate for multi-mode non-Gaussian operations can be orders of magnitude higher than single-mode operations.
Our results are important for real-world CV-QKD systems especially those dependent on quantum error correction - a process that requires non-Gaussian effects.

\end{abstract}

\section{Introduction}
CV-QKD protocols with non-Gaussian operations, such as photon subtraction and photon catalysis, have garnered great interest over the past few years (e.g. \cite{huang2013performance,hosseinidehaj2016cv,li2016non,guo2017performance,zhao2017improvement,zhao2018continuous,guo2019continuous,ye2019improvement,ye2019continuous}).
This is so partially because such operations are critical steps for quantum information tasks such as quantum error correction \cite{Nogo} and noiseless amplification \cite{jeffers2011optical,kim2012quantum,gagatsos2014heralded}, and that they have the potential to enhance the level of entanglement of entangled states such as Einstein-Podolsky-Rosen (EPR) states \cite{kitagawa2006entanglement,zhang2010distillation,navarrete2012enhancing,bartley2015directly,zhou2018entanglement}.
It is natural to hypothesize that non-Gaussian states can also boost the secret key rate of entanglement-based CV-QKD protocols.

Considering single-mode states, photon subtraction has been studied intensively in various entanglement-based CV-QKD protocols (e.g\cite{huang2013performance,li2016non,guo2017performance,zhao2017improvement,zhao2018continuous}).
Results of these studies show that photon subtraction can significantly increase the performance, such as the transmission distance and the excess noise tolerance of a CV-QKD system.
CV-QKD protocols with photon catalysis are also proposed in recent research (e.g., \cite{guo2019continuous,ye2019improvement,ye2019continuous}).
It turns out that, relative to photon subtraction, photon catalysis provides a higher secret key rate and a longer transmission distance.

All of the above performance improvements are under the assumption that the original entangled state (an EPR state) is strongly squeezed. A typical value for the variance of the EPR state adopted in previous works is 20, which corresponds to a 16dB squeezing.
However, for certain CV-QKD protocols, a more weakly-squeezed  ($\sim$12dB) ) EPR state leads to improved performance without performing any non-Gaussian operations  \cite{djordjevic2019photon,he2019photonic}.

However, the studies discussed above are all under the assumption that each beam of the EPR state only contains a single frequency mode.
In reality, any quantum state contains multiple frequency modes - an issue of increased concern when broadband pulses of light (narrow pulses in the time domain) are utilized.
The multi-mode entangled states potentially allow for a higher quantum channel capacity \cite{christ2012exponentially,usenko2014entanglement,hosseinidehaj2017multimode}.
It is therefore natural to investigate what impact non-Gaussian operations can have on a multi-mode CV-QKD system.

In this work, we aim to determine whether non-Gaussian operations can improve the performance of a multi-mode CV-QKD system.
The contributions of this paper are summarized as follows:
(i) We explore the performance of different multi-mode non-Gaussian operations at different loss values, showing that photon subtraction always outperforms photon addition.
(ii) We show that multi-mode non-Gaussian operations can significantly increase the optimized CV-QKD key rate.
(iii) We quantify that the improvement for multi-mode operations can be orders of magnitude higher than single-mode operations at losses of 30dB, which is the operational limit of current CV-QKD systems.

The structure of the remainder of this paper is as follows.
In Section \ref{sec:multimode}, the multi-mode entangled states and the multi-mode non-Gaussian operations are described.
In Section \ref{sec:protocol}, our multi-mode CV-QKD protocols with non-Gaussian operations are described, and in Section \ref{sec:result} our simulation results are presented.
	
\section{The Multi-mode States}\label{sec:multimode}
\subsection{The Multi-mode Entangled States}
The Parametric Down-Conversion (PDC) process is commonly used to create entangled states.
In reality, this process does not produce a single EPR state with single frequency modes, but an ensemble of independent EPR states with broadband frequency modes.
In the PDC process, a pump laser is first fed into a non-linear crystal. Two correlated beams, labeled as $\bm{A}$ and $\bm{B}$, are then created.
Each one of the beams contains multiple orthogonal broadband frequency modes, which are called supermodes.
Let $\hat{A}^\dagger_k$ and $\hat{B}^\dagger_k$ be the creation operator of the supermodes in beams $\bm{A}$ and $\bm{B}$, respectively, where the subscript $k\in\left\{1,2,...,\infty\right\}$ is used to index the supermodes. The output state of the PDC process can be written as \cite{christ2012exponentially}
\begin{equation}\label{eq:PDCinSch}
\begin{aligned}
\ket{\text{PDC}}_{AB} &= \bigotimes_{k=1}^\infty \exp\left[G \lambda_{k}\left(\hat{A}^\dagger_k \hat{B}^\dagger_k  - \hat{A}_k \hat{B}_k\right) \right]\ket{0}\\
&=\bigotimes_{k=1}^\infty \ket{\text{EPR}_k}_{AB},
\end{aligned}
\end{equation}
where
\begin{equation}\label{eq:EPRsingle}
\ket{\text{EPR}_k}_{AB}=\left(\sqrt{1-\tanh^2{r_k}}\right)\sum_{n=0}^\infty \tanh^n{r_k} \ket{n,n}_{AB},
\end{equation}
$r_k=G \lambda_{k}$ is the squeezing parameter, $G$ is the overall gain of the PDC process, and the $\lambda_{k}$'s are normalized coefficients, which follow an exponentially decaying distribution for the most likely PDC sources \cite{christ2012exponentially}.

The quadrature operators associated with one EPR state of the PDC state are defined as ($\hbar = 2$ is adopted)
\begin{align}
\begin{array}{l}{\hat{X}_{k}^{A}=\hat{A}_{k}+\hat{A}_{k}^{\dagger}, \hat{P}_{k}^{A}=i\left(\hat{A}_{k}^{\dagger}-\hat{A}_{k}\right),} \\ {\hat{X}_{k}^{B}=\hat{B}_{k}+\hat{B}_{k}^{\dagger}, \hat{P}_{k}^{B}=i\left(\hat{B}_{k}^{\dagger}-\hat{B}_{k}\right).}
\end{array}
\end{align}
Being an ensemble of Gaussian states, the PDC state can be fully characterized by the covariance matrix (CM) of the quadrature operators of the EPR states. The CM of each EPR state has the form
\begin{align}\label{eq:CMepr}
\Sigma_{k}=\left(\begin{array}{ll}
{\cosh \left(2 r_{k}\right) I} & {\sinh \left(2 r_{k}\right) Z} \\ 
{\sinh \left(2 r_{k}\right) Z} & {\cosh \left(2 r_{k}\right) I}
\end{array}\right),
\end{align}
where $I$ is the 2-by-2 identity matrix and $Z=\text{diag}[1,-1]$.

The quadrature operators of a multi-mode state can be measured by two different detection  strategies. The first strategy undertakes a multi-mode homodyne measurement (e.g., \cite{armstrong2012programmable, roslund2014wavelength,plick2018violating}), which applies a simultaneous measurement on all the supermodes. The second strategy, which allows for successive measurements on each supermode, first splits a supermode into different frequency bands using a Quantum Pulse Gate (QPG) (e.g., \cite{eckstein2011quantum, brecht2011quantum}), then undertakes a homodyne measurement to each of the split supermodes.

\subsection{Multi-mode States with Non-Gaussian Operations}
\begin{figure}
	\centering
	\includegraphics[width=0.98\linewidth]{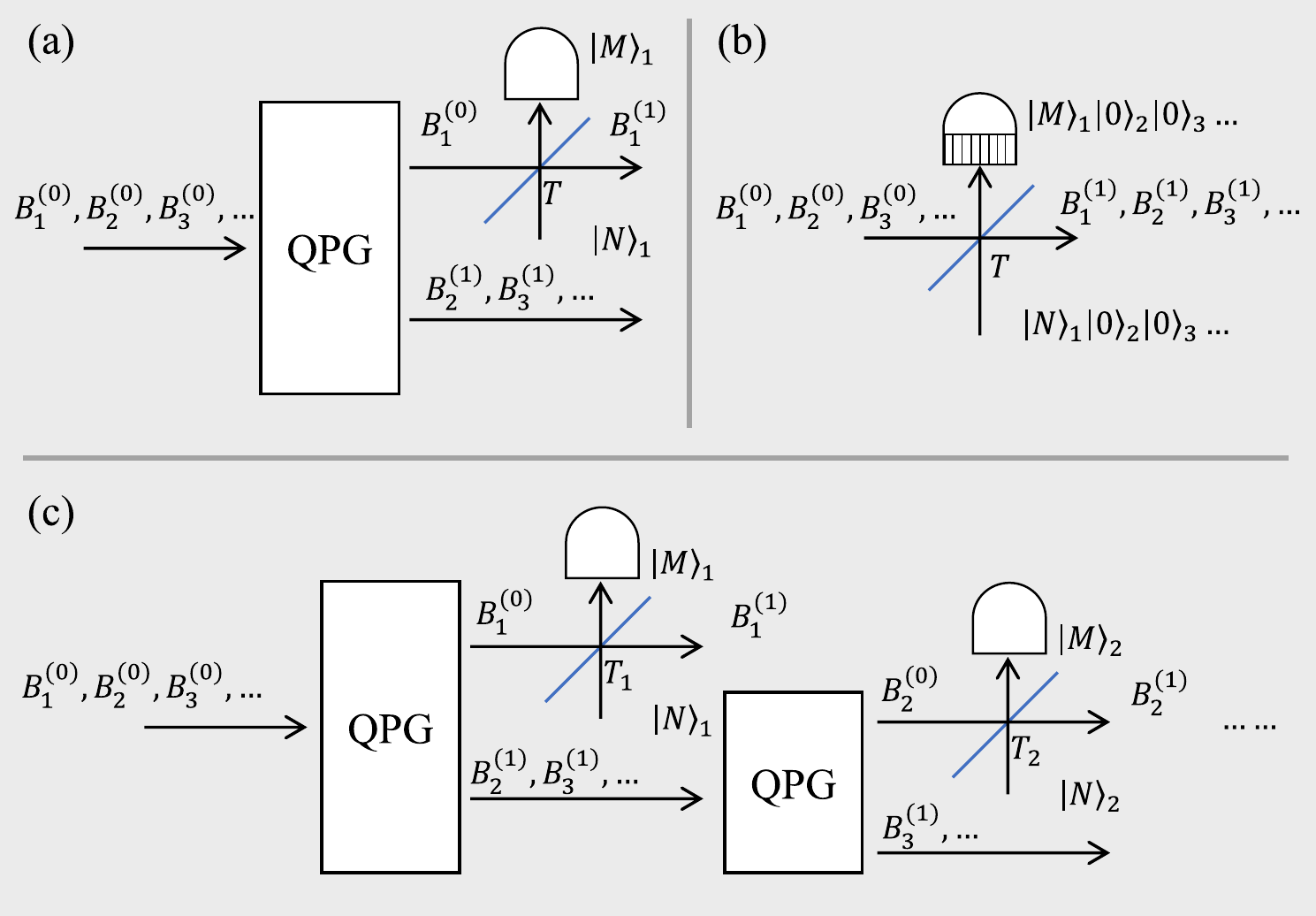}
	\caption{(a) The multi-mode non-Gaussian operation with a QPG. (b) The multi-mode non-Gaussian operation without a QPG. (c) A concatenation of multi-mode non-Gaussian operations with a QPG.}
	\label{fig:figdignongaussian}
\end{figure}

A typical experimental set-up for the multi-mode non-Gaussian operations is illustrated in Fig.~\ref{fig:figdignongaussian}a.
Suppose a non-Gaussian operation is to be performed to the first supermode of beam $\bm {B}^{(0)}$ of a PDC state, i.e., supermode $B^{(0)}_1$.
Note, for clarity we use the superscript $(\cdot)^{(n)} (n\in\{0,1,2...\})$ to distinguish a supermode (or a beam) at different stages of the process.
Beam $\bm {B}^{(0)}$ is first fed into a QPG. The gating pulse for the QPG is shaped to match the frequency profile of $B^{(0)}_1$, so that $B^{(0)}_1$ is selected among the supermodes.
The selected supermode then interacts with an ancillary $N$-photon state, $\ket{N}_1$, at a beam splitter with transmissivity $T$.
The frequency profile of the ancillary state is also shaped to match the profile of $B^{(0)}_1$.
The ancillary output is conditionally measured by a photon number detector, which only clicks for an $M$-photon output.
This conditional measurement will project $B^{(0)}_1$ to a non-Gaussian form.
After the non-Gaussian operation, the supermodes other than $B^{(0)}_1$ are left intact.

Depending on the ancillary state and the photon number detector, there are mainly three types of non-Gaussian operations.
These are the photon subtraction ($N=0$), the photon addition ($M=0$), and the photon catalysis ($N=M$).
In our previous work, we have shown that for a CV-QKD protocol, the single-photon non-Gaussian operations offer a higher
secret key rate and success probability relative to their multi-photon
 counterparts for a given type of non-Gaussian operation \cite{he2019photonic}.
As such, in this work we mainly focus on single-photon non-Gaussian operations.
We will also investigate the zero-photon non-Gaussian operations which we have not considered in our previous work.

The CM of the leading EPR state ($k=1$) is given by Eq.~(\ref{eq:CMepr}).
Performing the single-photon subtraction (1-PS) on $B^{(0)}_1$ will alter the CM to
\begin{align}\label{eq:cm1ps}
\Sigma_{1}^{(\text{1-PS})}=\left(\begin{array}{cc}
{\frac{3+\xi^2_1 T}{1-\xi^2_1 T} I} & {\frac{4\sqrt{\xi^2_1 T}}{1-\xi^2_1 T} Z} \\
{\frac{4\sqrt{\xi^2_1 T}}{1-\xi^2_1 T} Z} & {\frac{1 + 3\xi^2_1 T}{1-\xi^2_1 T} I}\end{array}\right),
\end{align}
where $\xi_1^2=\tanh^2 r_1$. The success probability for this operation is
\begin{equation}
P_1^{(\text{1-PS})}=\xi_1^2 (1-\xi_1^2)\frac{1-T}{\left(1-\xi_1^2T\right)^2}.
\end{equation}
Likewise, performing the single-photon addition (1-PA) will alter the CM to
\begin{align}\label{eq:cm1pa}
\Sigma_{1}^{(\text{1-PA})}=\left(\begin{array}{cc}
{\frac{1 + 3\xi^2_1 T}{1-\xi^2_1 T} I} & {\frac{4\sqrt{\xi^2_1 T}}{1-\xi^2_1 T} Z} \\
{\frac{4\sqrt{\xi^2_1 T}}{1-\xi^2_1 T} Z} & {\frac{3+\xi^2_1 T}{1-\xi^2_1 T} I}\end{array}\right).
\end{align}
The corresponding success probability for 1-PA is
\begin{equation}
P_1^{(\text{1-PA})}=\xi_1^2  \frac{1-T}{\left(1-\xi_1^2T\right)^2}.
\end{equation}
The CM for the resultant state after the single-photon catalysis (1-PC) is quite involved, and reads
\begin{align}\label{eq:cm1pc}
\Sigma_{1}^{(\text{1-PC})}=\left(\begin{array}{cc}
{a_1 I} & {c_1 Z} \\
{c_1 Z} & {b_1 I}\end{array}\right),
\end{align}
where
\begin{equation}
\begin{footnotesize}
\begin{aligned}
a_1&=\frac{T  - \xi_1^2 (-3 + 12 T - 8 T^2) - \xi_1^4 T (-8 + 12 T - 3 T^2)+ \xi_1^6 T^2}
{(-1 + \xi_1^2 T) [T  + \xi_1^2 (1 - 4 T + T^2)+ \xi_1^4 T]},\\
b_1&=a_1,\\
c_1&=\frac{2 \sqrt{T}\xi_1 [1 - 2 T - 2 \xi_1^2 (2 - 5 T + 2 T^2) + \xi_1^4 T (-2 + T)  ]}
{(-1 + \xi_1^2 T)[T  + \xi_1^2 (1 - 4 T + T^2)+ \xi_1^4 T]}.
\end{aligned}
\end{footnotesize}
\end{equation}
The success probability for 1-PC is
\begin{equation}
\begin{footnotesize}
\begin{aligned}
P_1^{(\text{1-PC})}=\frac{-T   + \xi_1^2 (-1 + 5 T - T^2) + \xi_1^4 (1 - 5 T + T^2)  + \xi_1^6 T}
{(-1 + \xi_1^2 T)^3}.
\end{aligned}
\end{footnotesize}
\end{equation}
Particularly, when $M=N=0$ we have the zero-photon catalysis (0-PC), which alters the CM to
\begin{align}\label{eq:cm0pc}
\Sigma_{1}^{(\text{0-PC})}=\left(\begin{array}{cc}
{\frac{1+\xi^2_1 T}{1-\xi^2_1 T} I} & {\frac{2\sqrt{\xi^2_1 T}}{1-\xi^2_1 T} Z} \\
{\frac{2\sqrt{\xi^2_1 T}}{1-\xi^2_1 T} Z} & {\frac{1+\xi^2_1 T}{1-\xi^2_1 T} I}\end{array}\right).
\end{align}
The success probability for 0-PC is
\begin{equation}
P_1^{(\text{0-PC})}=\frac{1-\xi_1^2}{1-\xi_1^2T}.
\end{equation}
Interestingly, the determinant of the CM in Eq.~(\ref{eq:cm0pc}) is equal to one, indicating that the 0-PC is a noiseless operation.

It is also possible to perform the non-Gaussian operations to more than one supermode.
Similar to the idea of the measurement of a multi-mode state, this can either be done by the method of using the multi-mode photon detection as is illustrated in Fig.~\ref{fig:figdignongaussian}b, or using a method based on the concatenation of the experimental set-up from Fig.~\ref{fig:figdignongaussian}a.
In the latter method, the values of the transmissivity of the beam splitter for different supermodes can be set individually.
For both methods the overall success probability of the non-Gaussian operations will be the product of the success probability for each operation.

\section{The Protocol for Multi-mode CV-QKD with Non-Gaussian Operations}\label{sec:protocol}
\begin{figure}
	\centering
	\includegraphics[width=0.98\linewidth]{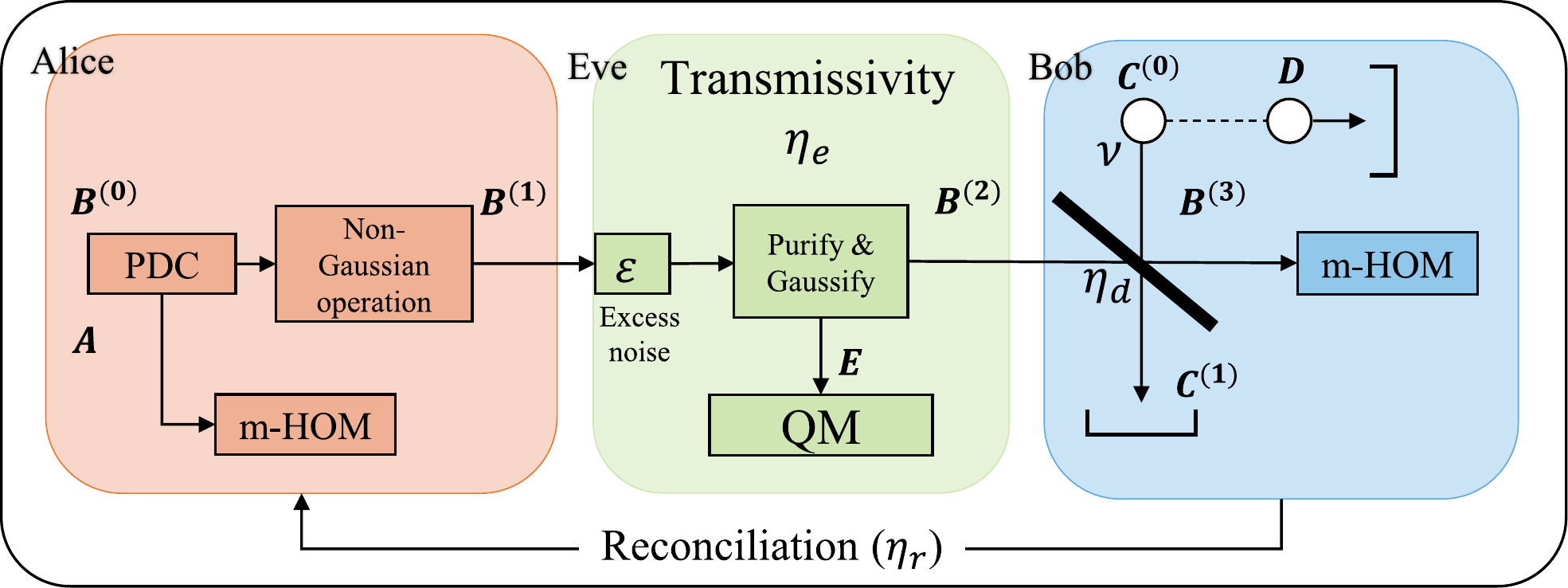}
	\caption{The protocol for multi-mode CV-QKD with non-Gaussian operations. QM: Quantum Memory, m-HOM, multi-mode homodyne detection.}
	\label{fig:figdigqkd}
\end{figure}

In this work, we consider the multi-mode version of an entanglement-based CV-QKD protocol with reverse reconciliation.
As illustrated in Fig.~\ref{fig:figdigqkd}, Alice first prepares her PDC state ($ \bm {A}- \bm {B}^{(0)}$).
For clarity, we assume that this state only has some finite number, $K_{\text{max}}$, of equivalent EPR states.
For the non-Gaussian operation we adopt the experimental set-up in Fig.~\ref{fig:figdignongaussian}c.
We assume Alice will apply the non-Gaussian operation to the first $K_{\text{sel}}$ supermodes, $B^{(0)}_1,B^{(0)}_2,...,B^{(0)}_{K_{\text{sel}}}$, of beam $\bm B^{(0)}$ of the PDC state.
Supermodes $B^{(0)}_{K_{\text{sel}}+1},...,B^{(0)}_{K_{\text{max}}}$ are left unchanged.
The resultant beam, which is labeled as $ \bm B^{(1)}$, is sent to Bob via a lossy and noisy channel, which is controlled by Eve.

The channel is characterized by the transmissivity $\eta_e$ and the  excess noise $\varepsilon$ (see appendix).
For the channel we have the following assumptions:
\begin{itemize}
	\item The channel loss is frequency independent, so that each supermode experiences the same level of attenuation.
	\item The excess noise is i.i.d. for each supermode.
	\item Eve has full knowledge of Alice and Bob's supermode structure and their non-Gaussian operation scheme.
\end{itemize}
Under our assumptions the supermode structure of beam $\bm B^{(1)}$ is retained after the channel. The multi-mode channel is equivalent to multiple independent single-supermode sub-channels.
For each sub-channel, we assume Eve will first Gaussify the incoming supermode $B_{k}^{(1)}$ if a non-Gaussian operation has been performed to this supermode (i.e., she will replace the non-Gaussian state $\rho_{AB^{(1)}}$ with a Gaussian state that has the same CM and frequency profile as that of $\rho_{AB^{(1)}}$).
Eve will then perform an entanglement cloning attack to obtain a purification of $B_{k}^{(1)}$.
Eve will hide herself by mimicking the anticipated noise conditions.
Eve will then store her ensemble of purification, $\bm E$, in her quantum memory and will perform a joint measurement on $\bm E$ after the reverse reconciliation process.

The beam after the channel is labeled as $ \bm B^{(2)}$.
Bob will perform an imperfect multi-mode homodyne measurement characterized by a detection efficiency $\eta_d$ and a thermal noise $\nu$ on the received beam.
We assume that the thermal noise is i.i.d. among the sub-channels.
For each sub-channel, the detection efficiency is modeled by a beam splitter with transmissivity $\eta_d$, while the thermal noise is modeled by an EPR state ($  C^{(0)}_k-  D_k$) with variance given by $\nu$.
The frequency profile of supermode $C^{(0)}_k$ is identical to that of $B^{(2)}_k$.
Supermode $B^{(2)}_k$ first interacts to $C^{(0)}_k$ at the beam splitter.
We assume Eve does not have access to Bob's detector, so that the split supermode, $C^{(1)}_k$, and supermode $D_k$ are discarded after the interaction.
The ensemble of the output supermodes, labeled as $\bm B^{(3)}$, is then injected into a perfect multi-mode homodyne detector.

\begin{figure*}[tb]
	\centering
	\begin{subfigure}{
			\includegraphics[width=.40\linewidth]{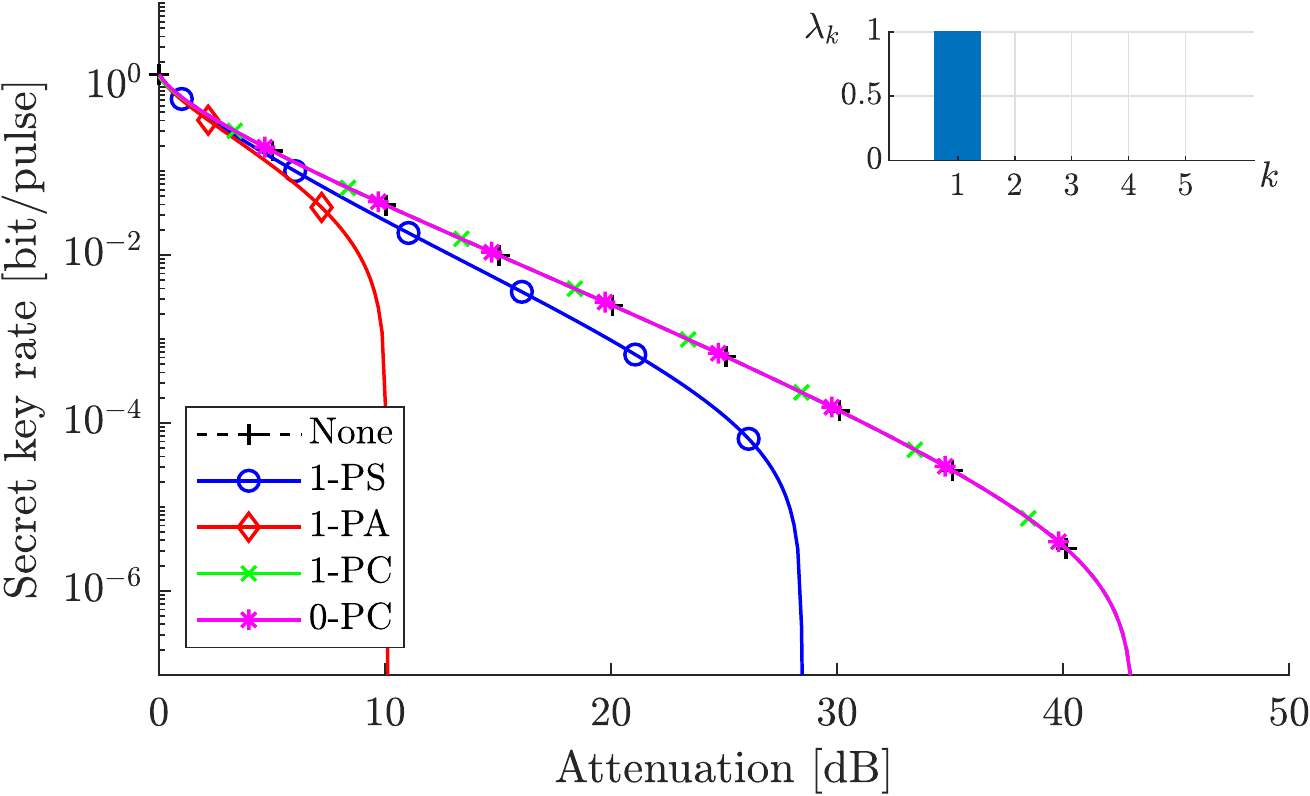}}
	\end{subfigure}\qquad
	\begin{subfigure}{
			\includegraphics[width=.40\linewidth]{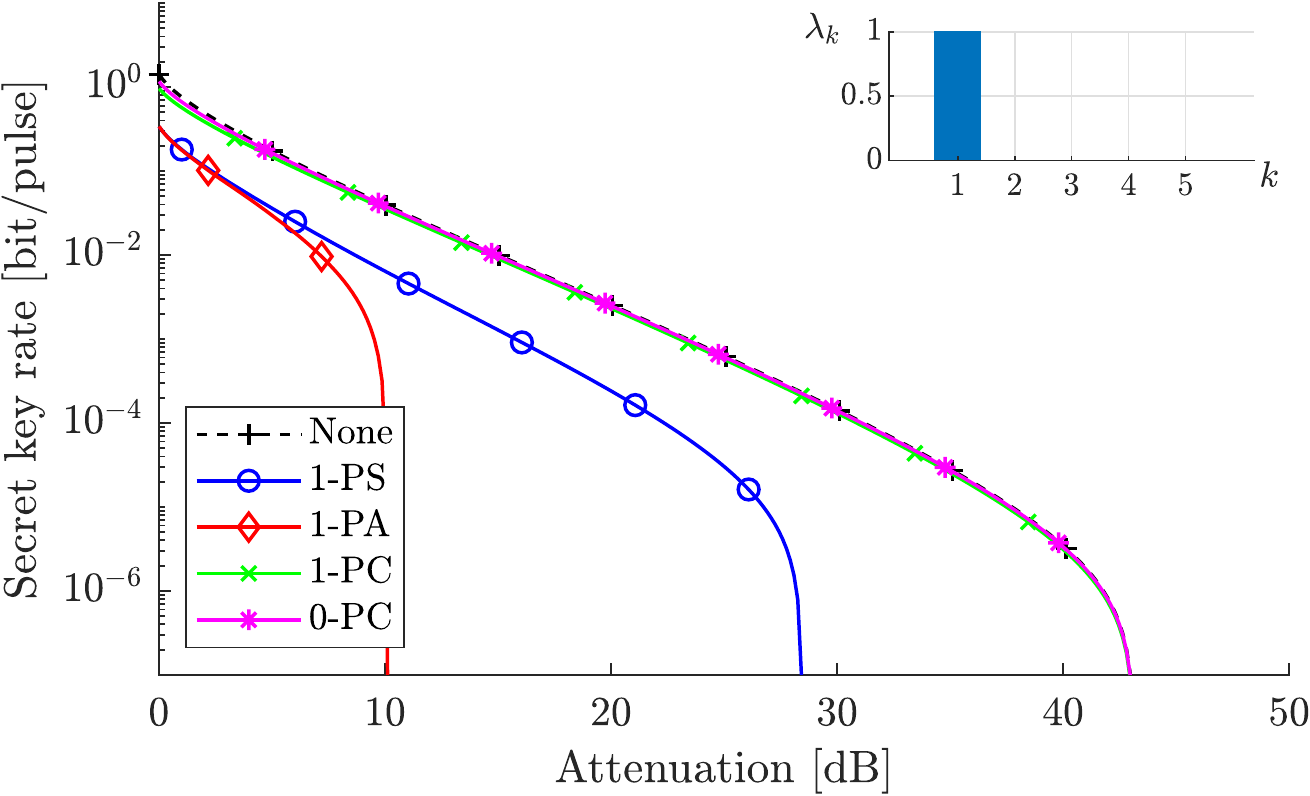}}
	\end{subfigure}	
	\\
	\begin{subfigure}{
			\includegraphics[width=.40\linewidth]{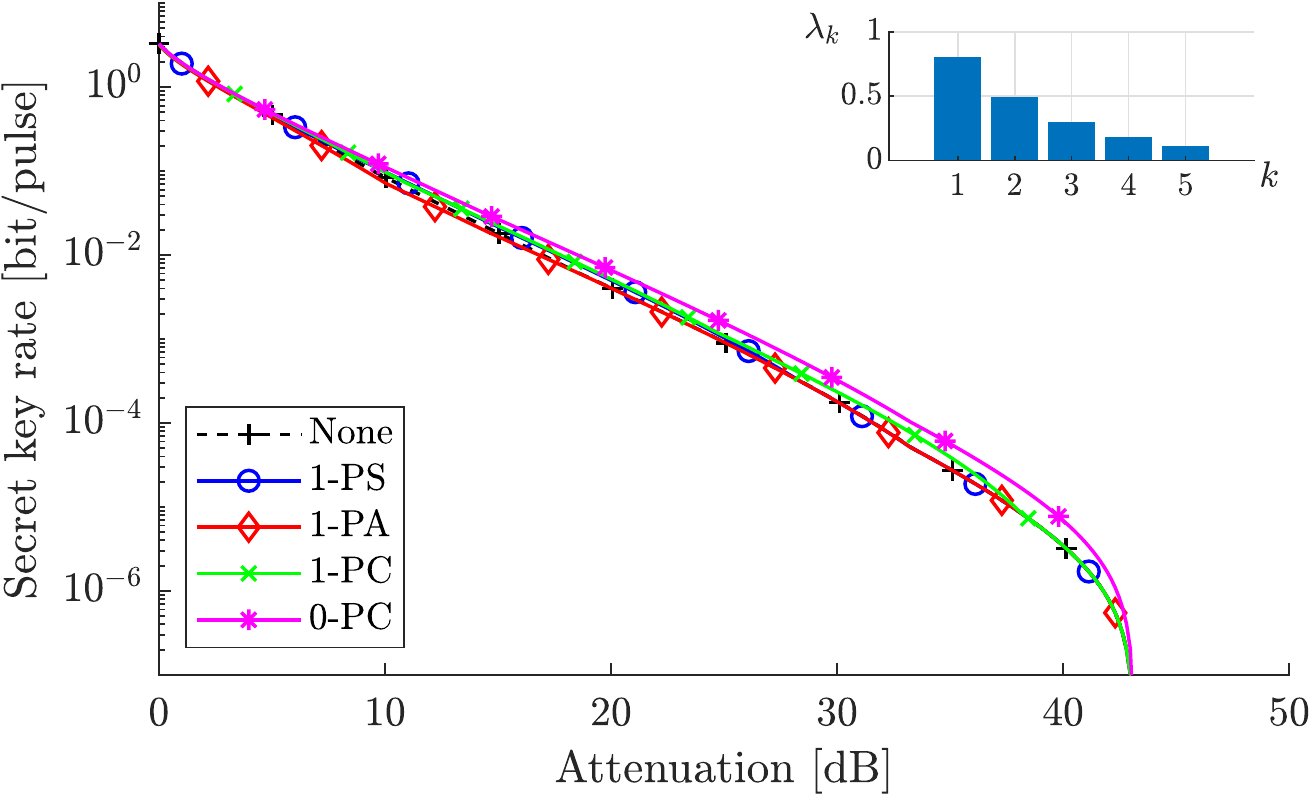}}
	\end{subfigure}\qquad
	\begin{subfigure}{
			\includegraphics[width=.40\linewidth]{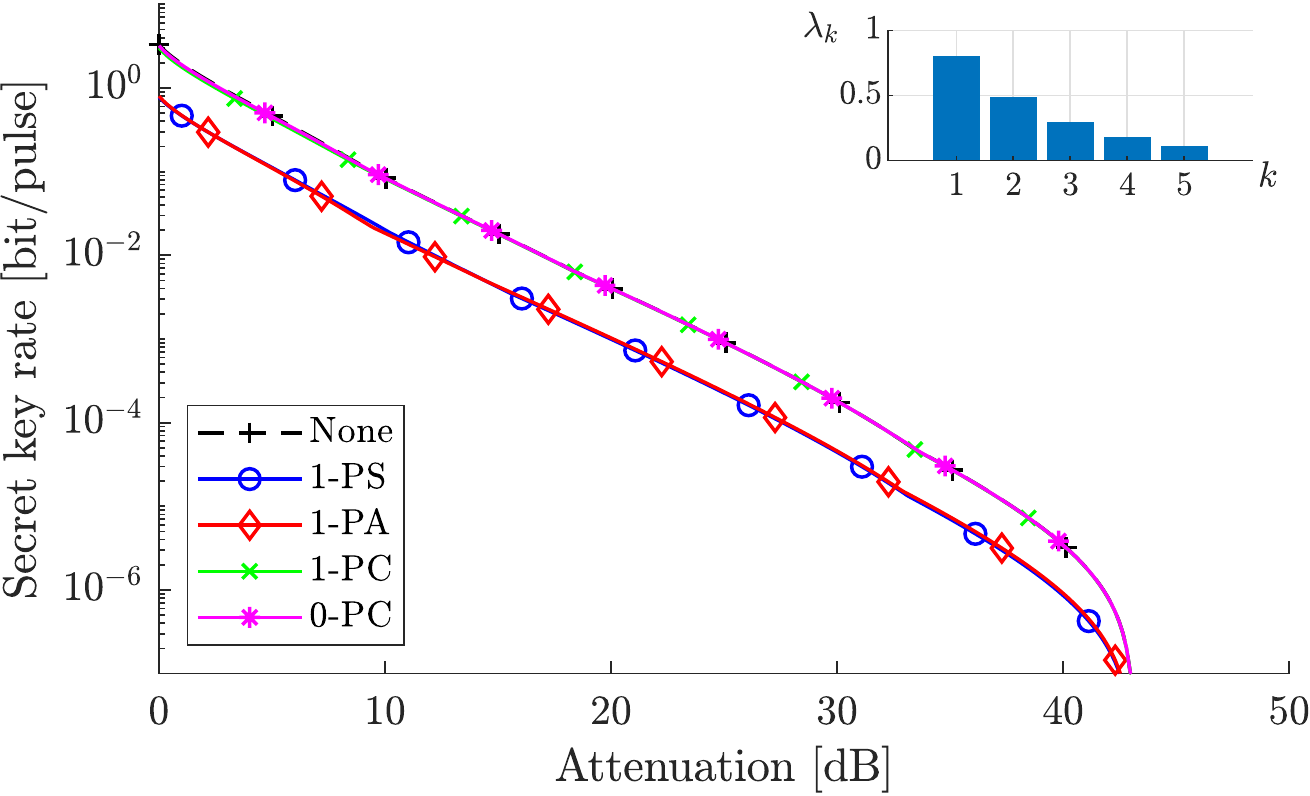}}
	\end{subfigure}
	\\
	\begin{subfigure}{
			\includegraphics[width=.40\linewidth]{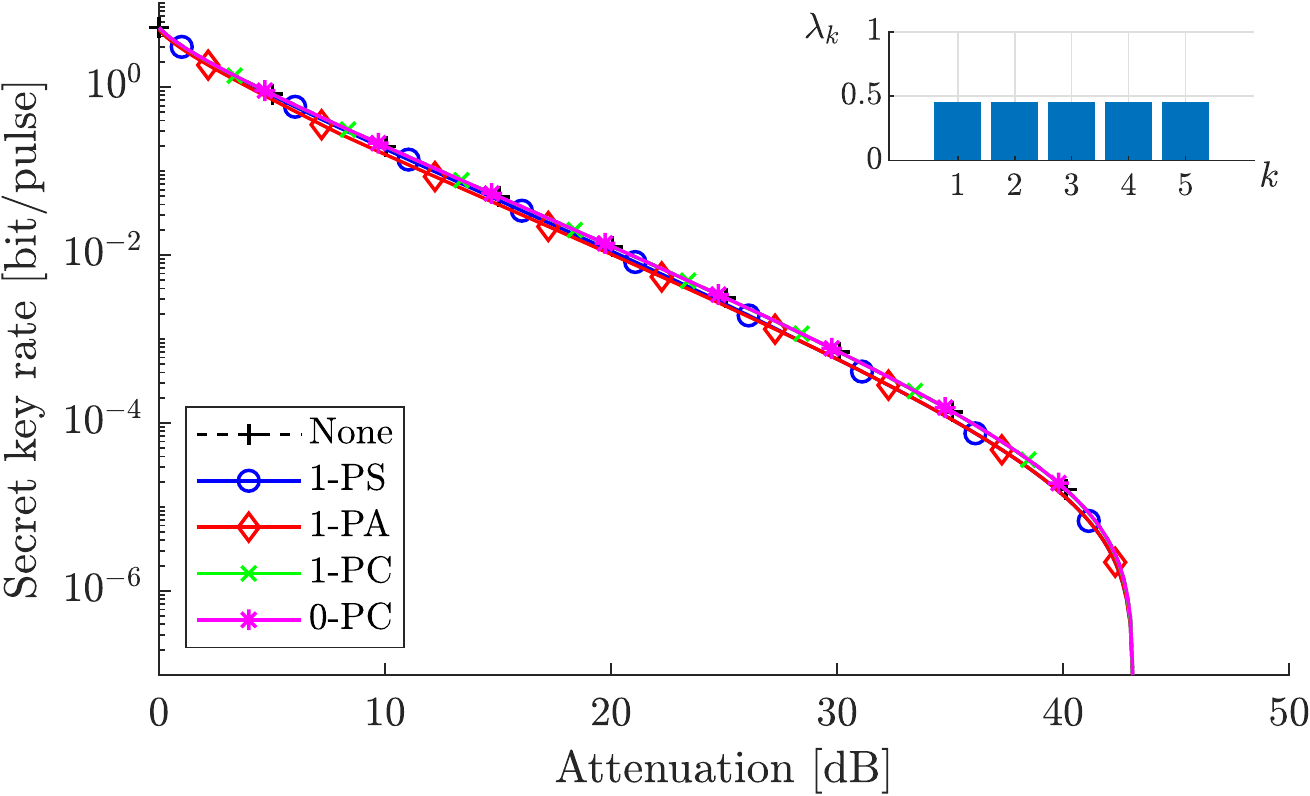}}
	\end{subfigure}\qquad
	\begin{subfigure}{
			\includegraphics[width=.40\linewidth]{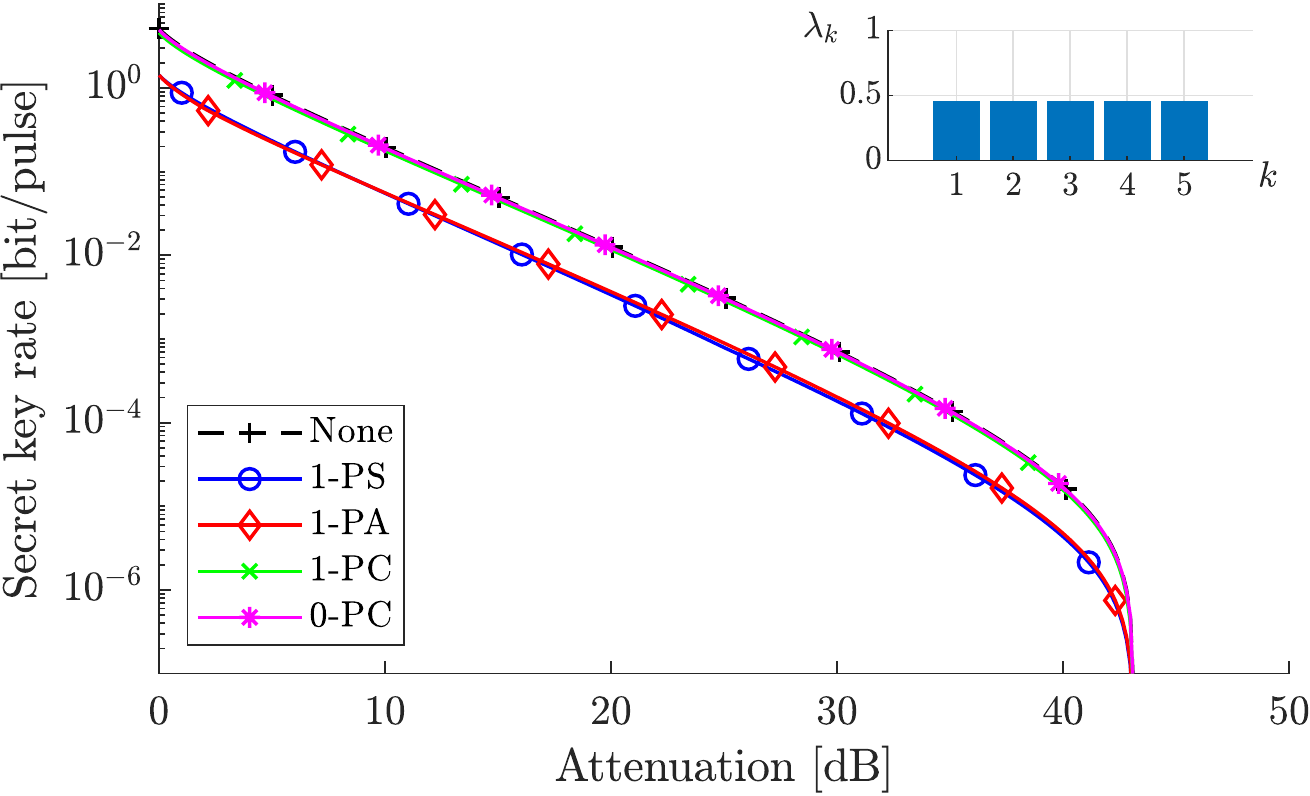}}
	\end{subfigure}
	\caption{Figures in the the left panel illustrate the optimized secret key rate with quantum memory devices.
		The insets of each figure illustrate the supermode structure of the PDC states.
		Figures in the right panel illustrate the optimized secret key rate without quantum memory devices. The non-Gaussian operation is only applied to the first supermode.
	}
	\label{fig:1nonG}
\end{figure*}

We assume a quantum memory device is available at Alice's side, so that she can prepare the non-Gaussian state in advance. In this case, the success probability for the non-Gaussian operation can be viewed as one.
Under the assumption of infinite key size, the secret key rate for our multi-mode non-Gaussian CV-QKD protocol is lower bounded by \cite{navascues2006optimality}
\begin{equation}\label{keyrate}
R_{\text{tot}} \geq \sum_{k=1}^{K_{\text{max}}}R_k,
\end{equation}
where
\begin{equation}
R_k = \eta_r I(A_k\negmedspace:\!B_k^{(3)}) - \chi(E_k\negmedspace:\!B_k^{(3)})
\end{equation}
is the secret key rate for each sub-channel, $\eta_r$ is the reverse reconciliation efficiency, $I(A_k\negmedspace:\!B_k^{(3)})$ is the classical mutual information between Alice and Bob, and $\chi(E_k\negmedspace:\!B_k^{(3)})$ is the Holevo bound for Eve's information.

Alice and Bob's mutual information can be calculated by
\begin{equation}
I(A_k\negmedspace:\!B_k^{(3)})=\frac{1}{2}\log_2{\frac{V_{A_k}}{V_{A_k|B_k^{(3)}}}}\textrm{,}
\label{AandB}
\end{equation}
where $V_{A_k}$ is the variance of Alice's supermode, and ${V_{A_k|B_k^{(3)}}}$ is the variance of Alice's supermode conditioned on Bob's homodyne measurement.
The Holevo bound for Eve's information is
\begin{equation}
\chi(E_k\negmedspace:\!B_k^{(3)}) =\sum_{i=1}^{2} g\left( \alpha_{i,k}\right) - \sum_{j=3}^{5} g\left( \alpha_{j,k}\right)\textrm{,}
\label{EandB}
\end{equation}
where
$g(x)=\frac{x+1}{2}\log_2\frac{x+1}{2}-\frac{x-1}{2}\log_2\frac{x-1}{2}$, $\alpha_{i,k}$'s are the symplectic eigenvalues of the CM of  state $\rho_{AB^{(2)}}$, and $\alpha_{j,k}$'s are the symplectic eigenvalues of the CM of  state $\rho_{AC^{(1)}D}$ conditioned on Bob's measurement.

If a quantum memory device is not available to Alice, the secret key rate is lower bounded by
\begin{equation}\label{keyratenomem}
R_{\text{tot}} \geq \left(\prod_{k=1}^{K_{\text{sel}}} P_k\right) \sum_{k=1}^{K_{\text{max}}}R_k,
\end{equation}
where $P_k$ is the success probability for the non-Gaussian operation for each sub-channel.

For calculation of the secret key rate we need to calculate $I(A_k\negmedspace:\!B_k^{(3)})$ and $\chi(E_k\negmedspace:\!B_k^{(3)})$, which are determined by the CMs of the states $\rho_{AB^{(2)}}$ and $\rho_{AC^{(1)}D}$, respectively. The derivation of the two CMs is straightforward but lengthy. Readers can find details for the derivation in the appendix.

\section{Simulation Results}\label{sec:result}
For each sub-channel, we adopt the same noise and efficiency parameters from our previous work \cite{he2019photonic}.
Specifically, we set the excess noise $\varepsilon=0.1$, the detection noise $\nu=1.1$ (both in vacuum noise units), the detection efficiency $\eta_d=0.68$, and the reconciliation efficiency $\eta_r=0.95$.
We assume the number of equivalent EPR states of the PDC state created by Alice is $K_{\text{max}}=5$.
These EPR states are characterized by the squeezing parameters $\left[r_1,r_2,...,r_5\right]=G\left[\lambda_1,\lambda_2,...,\lambda_5\right]$.
For the normalized coefficients of the PDC state $\lambda_1,\lambda_2,...,\lambda_5$, we consider three scenarios.
In the first scenario the coefficients are all zero except $\lambda_1$. This state is a good approximation to a single-mode state.
In the second scenario the coefficients follow an exponentially decaying distribution of coefficient values (i.e squeezing), which is the most likely case expected in real PDC sources. 
We refer to this as a \emph{generic} supermode system.
In the last scenario the coefficients are all identical.

We first investigate the strategy where Alice only performs the non-Gaussian operation to the first supermode.
For each channel attenuation level, we assume Alice will adjust the value for the overall gain for the PDC process, $G$, and the transmissivity for the non-Gaussian operation, $T$, to obtain the \emph{optimized secret key rate}.
The results are illustrated in Fig.~\ref{fig:1nonG}.
The photon catalysis cannot improve the optimized secret key rate or transmission distance for a single-mode CV-QKD system.
However, as is shown in the middle figures, for the most likely PDC source, applying the photon catalysis or the photon subtraction to the leading supermode can increase the optimized secret key rate when the quantum memory devices are available.
We can also see from the bottom figures that none of the three types of non-Gaussian operation can improve the secret key rate or transmission distance when the squeezing of the PDC state is evenly distributed among the supermodes, even if the quantum memory devices are available to Alice.

\begin{figure}[tb]
	\centering
	\begin{subfigure}{
		\includegraphics[width=.8\linewidth]{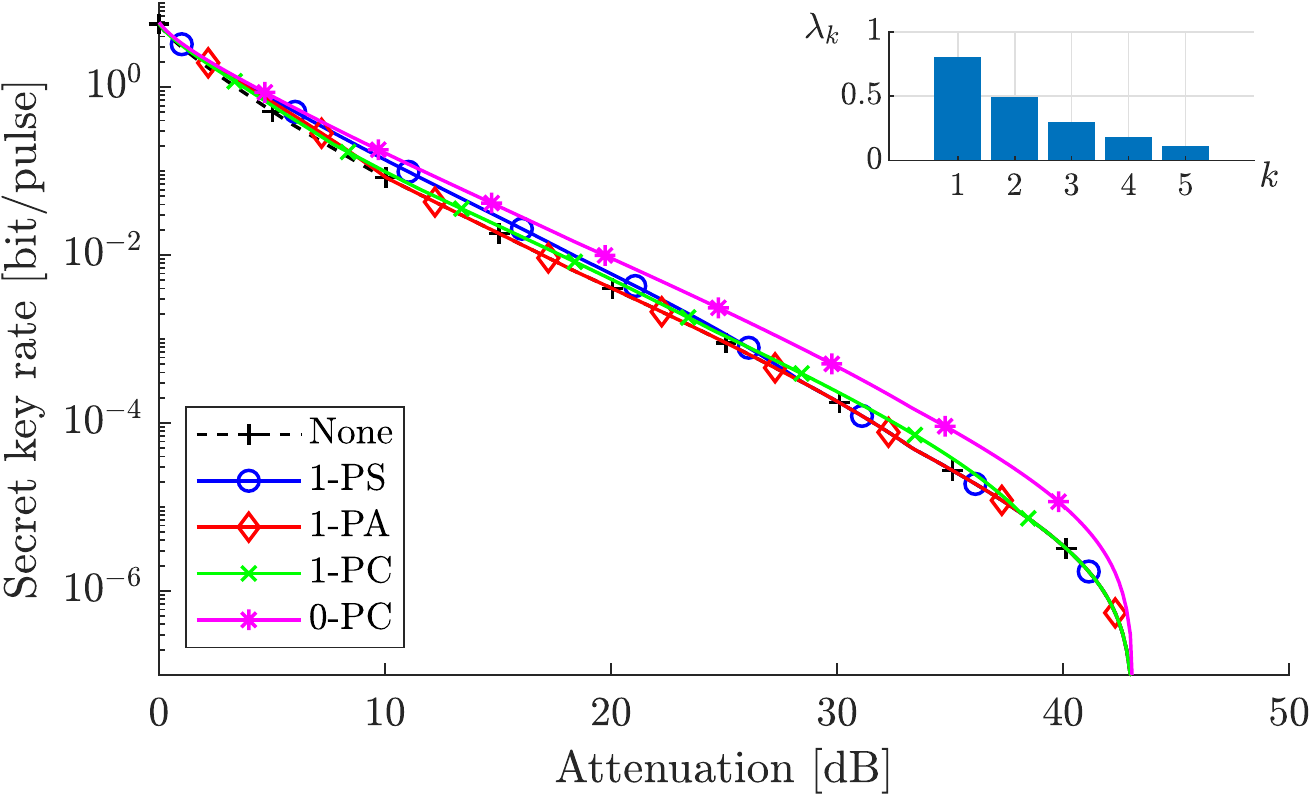}}
	\end{subfigure}
	\begin{subfigure}{
		\includegraphics[width=.8\linewidth]{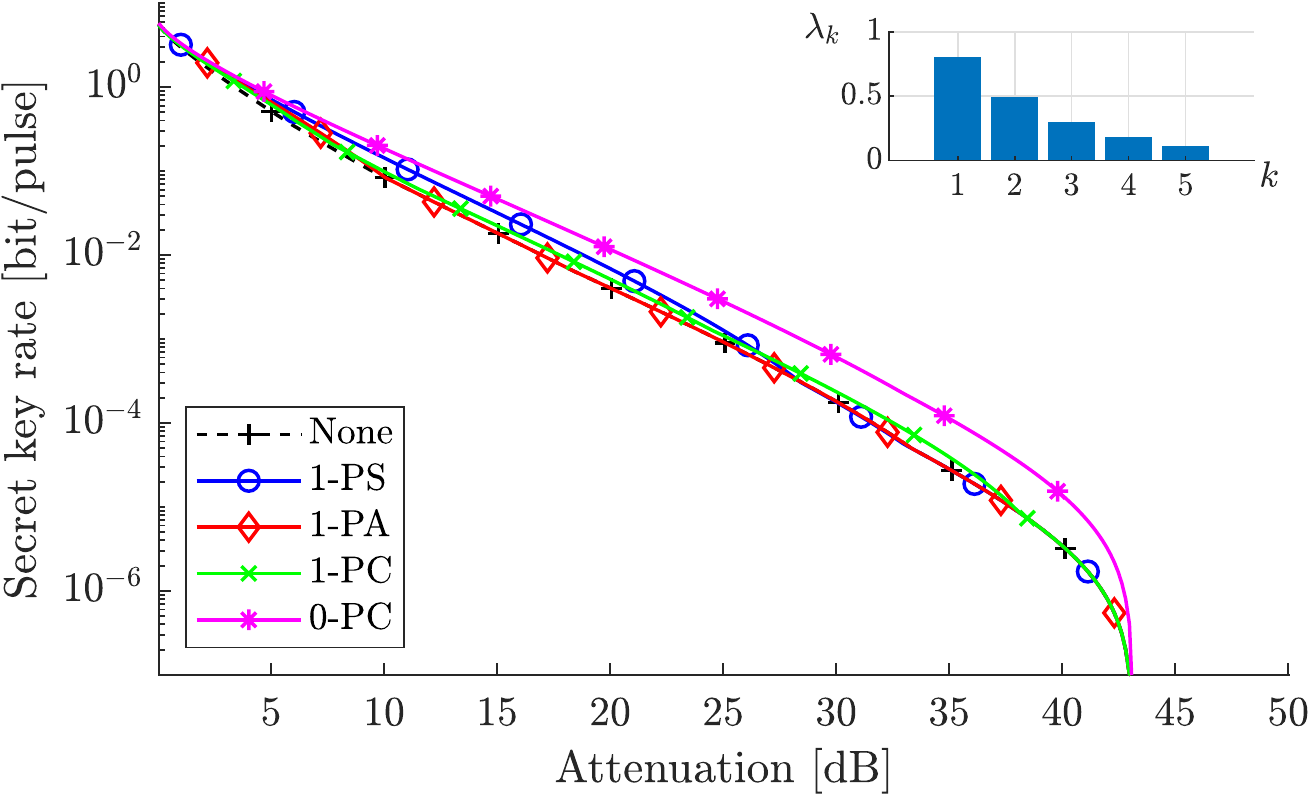}}
	\end{subfigure}
	\caption{The optimal secret key with the non-Gaussian operation applied to the first two (top) or three (bottom) supermodes.}
	\label{fig:3nonG}
\end{figure}

We then investigate the strategy where Alice performs the non-Gaussian operation to more than one supermode.
Specifically, we assume Alice performs the same type of non-Gaussian operation to the first two supermodes ($K_{\text{sel}}=2$).
The gain for the PDC process, $G$, and the values of the transmissivity for the two successive non-Gaussian operations, $T_1$ and $T_2$, are adjusted for the optimized secret key rate.
In this scheme we only consider the PDC state with exponentially decaying coefficients.
We again assume the quantum memory devices are available to Alice.
For comparison we also investigate the scheme where the non-Gaussian operation is applied to the first three supermodes.
The results are shown in Fig.~\ref{fig:3nonG}.
All the three types of non-Gaussian operations can increase the optimized secret key rate.
Among these non-Gaussian operations the zero-photon catalysis shows the best improvement.


The main conclusion from the above results is the following. Contrary to single-mode CV-QKD systems, in generic multi-mode CV-QKD systems it is possible to use non-Gaussian operations to increase the optimized secret key rate.
We attempt to better explain this result. In optimizing the key rate for single-mode systems, we optimize the squeezing of the initial entangled state sent through the channel, based on the loss of that channel. Our numerical results show that non-Gaussian operations can never improve upon this optimization strategy. For our multi-mode CV-QKD system the supermodes are equivalent to multiple single-mode channels.
For a generic supermode system that follow an exponentially decaying distribution (in squeezing), when we adjust the squeezing of the supermodes, we can only optimize the squeezing for one supermode. It is this fact that allows the non-Gaussian operations to have a different effect in terms of key rate, relative to single modes. In the rare occurrence where the supermodes possess the same squeezing levels, the enhanced key rates produced by non-Gaussian operations would not be present.

\section{Conclusion}
We have investigated the use of three types of non-Gaussian, namely the photon subtraction, the photon addition, and the photon catalysis, in the context of multi-mode entanglement-based CV-QKD.
We have shown that while these non-Gaussian operations cannot increase the optimized secret key rate for a single-mode CV-QKD system, for a generic multi-mode CV-QKD system an increase is possible by applying the non-Gaussian operations.
Among the non-Gaussian operations we have investigated, the zero-photon catalysis provides the highest improvement.


\section*{Appendix: Derivation of the Covariance Matrix}
Noticing the evolution of the CM follows the same procedure for every EPR state, we use the first EPR state as an example to derive the CMs.
Let
\begin{align}
\Sigma_{A_1B_1^{(1)}}=\left(\begin{array}{cc}
{a I} & {c Z} \\
{c Z} & {b I}\end{array}\right),
\end{align}
be the CM of the EPR state after the non-Gaussian operation (which is selected from Eqs.~(\ref{eq:CMepr}), (\ref{eq:cm1ps}), (\ref{eq:cm1pa}), (\ref{eq:cm1pc}), and (\ref{eq:cm0pc}), depending on the non-Gaussian operation scheme).
After the channel, the above CM evolves to
\begin{align}\label{eq:cmeig1}
\Sigma_{A_1B_1^{(2)}} &= \left[ {\begin{array}{*{20}{cc}}
	aI &\sqrt{\eta_e}cZ \\
	\sqrt{\eta_e}cZ&\left[\eta_e(b+\varepsilon)+(1-\eta_e)\right] I  \\
	\end{array}}\right]\nonumber\\
&:=\left[ {\begin{array}{*{20}{cc}}
	a I & c'Z \\
	c'Z & b'I  \\
	\end{array}}\right].
\end{align}
The CM of the state ($\rho_{A_1B_1^{(3)}C_1^{(1)}D_1}$) before Bob's homodyne detection is
\begin{align}
\Sigma_{A_1 B_1^{(3)} C_1^{(1)} D_1} &= \left[ {\begin{array}{*{20}{cc}}
	\Sigma_{A_1 C_1^{(1)} D_1} &\Gamma_{A_1 B_1^{(3)} C_1^{(1)} D_1} \\
	\Gamma_{A_1 B_1^{(3)} C_1^{(1)} D_1}^T&\Sigma_{B_1^{(3)}}\\
	\end{array}}\right]\textrm{,}
\end{align}
where the sub-matrices are defined as
\begin{equation}
\begin{small}
\begin{aligned}
&\Sigma_{A_1 C_1^{(1)} D_1} = \\
&\left[ {\begin{array}{*{20}{cc}}
	aI&-\sqrt{1-\eta_d}c'Z&0\\
	-\sqrt{1-\eta_d}c'Z&\left[\eta_d \nu+(1-\eta_d)b'\right]I&\sqrt{\eta_d(\nu^2-1)}Z\\
	0&\sqrt{\eta_d(\nu^2-1)}Z&\nu I
	\end{array}}\right],
\end{aligned}
\end{small}
\end{equation}
\begin{equation}
\Gamma_{A_1 B_1^{(3)} C_1^{(1)} D_1}=\left[\begin{array}{*{20}{cc}}
\sqrt{\eta_d}c'Z\\
\sqrt{(1-\eta_d)\eta_d}(\nu-b')Z\\
\sqrt{(1-\eta_d)(\nu^2-1)} Z
\end{array}\right]\textrm{,}
\end{equation}
and
\begin{equation}
\Sigma_{B_1^{(3)}}=\left[\eta_d b'+(1-\eta_d)\nu\right]I=:b''I.
\end{equation}
The CM of the state $\rho_{A_1 C_1^{(1)}D_1}$ conditioned on Bob's homodyne measurement is
\begin{equation}\label{eq:cmeig2}
\begin{small}
\begin{aligned}
&\Sigma_{A_1 C_1^{(1)}D_1|B_1^{(3)}}=\\
&\qquad\Sigma_{A_1 C_1^{(1)} D_1}-\frac{1}{b''}\Gamma_{A_1 B_1^{(3)} C_1^{(1)} D_1}X\Gamma_{A_1 B_1^{(3)} C_1^{(1)} D_1}^T\textrm{,}
\end{aligned}
\end{small}
\end{equation}
where $X=\textrm{diag}(1,0)$ for a $x$-quadrature measurement and $X=\textrm{diag}(0,1)$ for a $p$-quadrature measurement.
The mutual information in Eq.~(\ref{AandB}) can now be expressed as
\begin{equation}\label{AandBre}
\begin{small}
\begin{aligned}
&I(A_1\negmedspace:\!B_1^{(3)})=\\
&-\frac{1}{2}\log_2{\left\{{1-\frac{\eta_d \eta_e  c^2}{\eta_d \eta_e  ab+\left[\eta_d\left[(1-\eta_e)+\eta_e \varepsilon \right] + (1-\eta_d)\nu\right]a}}\right\}}\ \textrm{,}
\end{aligned}
\end{small}
\end{equation}
Let $\alpha_{1,1}$ and $\alpha_{2,1}$ be the symplectic eigenvalues of the CM in Eq.~(\ref{eq:cmeig1}), and $\alpha_{3,1}$, $\alpha_{4,1}$, and $\alpha_{5,1}$ be the symplectic eigenvalues of the CM in Eq.~(\ref{eq:cmeig2}), the Holevo bound for Eve's information can be calculated by putting symplectic eigenvalues  $\alpha_{1,1}$ to $\alpha_{5,1}$ into Eq.~(\ref{EandB}).



\end{document}